\documentclass[aps,prl,twocolumn,groupedaddress,showpacs]{revtex4}
\usepackage{graphicx}% Include figure files
\usepackage[dvips]{epsfig}% Include figure files
\usepackage{dcolumn}% Align table columns on decimal point
\usepackage{subfigure}%
\usepackage{bm}% bold math
\usepackage{amssymb}
\usepackage{amsmath}

\begin{document}

\title{Addendum to the Reply Comment [Phys. Rev. Lett. {\bf 102},
139602 (2009)]}

\author{Carlos Escudero}

\affiliation{Instituto de Ciencias Matem\'{a}ticas, Consejo Superior
de Investigaciones Cient\'{\i}ficas, C/ Serrano 123, 28006 Madrid,
Spain}

\begin{abstract}

\end{abstract}

\pacs{68.35.Ct,02.40.Ky,05.40.-a,68.35.Fx}

\maketitle

The object of this Addendum is to point out an important omission
in Reply \cite{escuderor} to Comment \cite{krug}. The omitted
argument, presented here, proves the incorrectness of several
arguments in the Comment. Comment \cite{krug} discusses the
connection of the radial correlations computed in \cite{escudero}
with the Family-Vicsek ansatz that describes the scaling behavior
of planar stochastic growth equations. According to it, these
correlations are a direct consequence of the mentioned ansatz,
this is, the two-points correlation function is of the
Family-Vicsek form
\begin{equation}
C(\theta,\theta',t) \approx t^{2\beta}
\mathcal{C}(|\theta-\theta'|t^{1-1/z}),
\end{equation}
where $\theta$, $\theta'$ represent sets of angles parameterizing
the $d-$dimensional radial interface. Then, the long time limit
{\it as taken in} \cite{krug} yields
\begin{equation}
\label{longtime}
\lim_{t \to \infty} C(\theta,\theta',t) \sim
t^{2\beta -1 +1/z}\delta(\theta - \theta')= t^{(1-d)/z}
\delta(\theta - \theta'),
\end{equation}
{\it according to} \cite{krug}, where the relation $2 \beta = 1-
d/z$ for linear growth equations has been employed in the last
equality. For $d=1$ we found a prefactor $t^0$, what is claimed in
\cite{krug} to be the explanation of the logarithmic prefactor in
\cite{escudero}. This argument is based on two erroneous facts.
First, the long time limit (\ref{longtime}) is incorrect, as the
prefactor necessary to build the Dirac delta function has been
forgotten. The correct calculation would be
\begin{eqnarray}
\label{longtime2} &C(\theta,\theta',t)& \approx t^{2\beta}
\mathcal{C}(|\theta-\theta'|t^{1-1/z})= \\
&t^{2\beta} t^{d/z-d}& \left[ \frac{1}{t^{d/z-d}}
\mathcal{C}(|\theta-\theta'|t^{1-1/z}) \right] \sim t^{1-d}
\delta(\theta - \theta'), \nonumber
\end{eqnarray}
where the long time limit has been taken in the last step.
Prefactors in (\ref{longtime}) and (\ref{longtime2}) are
different, and it is evident that result (\ref{longtime}) is
incorrect for two reasons. First, (\ref{longtime}) is not
compatible with the random deposition correlation $C_{rd} \sim t
\delta(x-x')$, for arc-length scales $x-x'=t(\theta-\theta')$,
obtained from the Family-Vicsek ansatz, and it does not coincide
with the radial correlations found for $d>1$ in \cite{escudero}
and \cite{escudero2}. Derivation (\ref{longtime2}) is compatible
with the random deposition correlation and with previously
calculated radial correlations {\it if dilution is taken into
account} (the term {\it dilution} refers to matter redistribution
due to substrate growth, see \cite{escudero3}). And this is
precisely the second mistake in \cite{krug}. Radial correlations
for which dilution has not been considered do not reduce to either
the incorrect (\ref{longtime}) or correct (\ref{longtime2}) forms.
One can see that, for $d>1$, the prefactor becomes constant and
not a power law of time \cite{escudero,escudero2}. And so, {\it
radial correlations cannot be deduced from the Family-Vicsek
ansatz}. On the other hand, if one takes into account dilution
\cite{escudero3}, radial correlations reduce to the Family-Vicsek
form (\ref{longtime2}) (but not to (\ref{longtime})). The physical
reason is that the radial stochastic growth equations considered
in all previous works (except those including dilution in
\cite{escudero3}) develop memory with respect to the initial
condition, and this memory effect is not captured by the
Family-Vicsek ansatz. Dilution erases this memory, and so its
inclusion implies the recovery of the Family-Vicsek ansatz
\cite{escudero3}. A physical argument also explains the
incorrectness of correlation (\ref{longtime}). The appearance of a
Dirac delta correlation for long times and large spatial scales
implies the uncorrelated character of the interface. The loss of
correlation is due to the fast domain growth, which rends
diffusion inoperative in this limit \cite{escudero3}. This implies
that the surface two-points correlation function must be
independent of the diffusion mechanism for large scales,
including, of course, the diffusion constant but also the order of
the linear diffusion operator $z$. Note that the incorrect
(\ref{longtime}) depends on $z$ but the correct (\ref{longtime2})
does not. Incorrect form (\ref{longtime}) expresses a
contradiction: the prefactor signals the measurable effect of
diffusion while the Dirac delta states that diffusion is
inoperative in this limit. As a side note let us mention that the
claim in \cite{krug} specifying that a prefactor $t^0$ in
(\ref{longtime}) (or (\ref{longtime2})) is compatible with the
logarithmic prefactor in \cite{escudero} is incorrect too.
Although these prefactors are compatible from a dimensional
analysis viewpoint, the explicit calculation of the two-points
correlation function shows that this prefactor is constant and not
logarithmic when $d=1$ \cite{escudero3}. Again, the logarithmic
prefactor is a memory effect that cannot be explained using the
Family-Vicsek ansatz.

In summary, the radial correlations calculated in \cite{escudero}
cannot be deduced from the Family-Vicsek ansatz, what implies a
different type of scaling. The arguments presented here reinforce
the conclusions of both Letter \cite{escudero} and Reply
\cite{escuderor} about the possible necessity of reconsidering
those experimental works in which radial interface profiles were
analyzed applying planar concepts without any justification.


\begin{thebibliography} {99}

\bibitem{escuderor} C. Escudero, Phys. Rev. Lett. {\bf 102},
139602 (2009).

\bibitem{krug} J. Krug, Phys. Rev. Lett. {\bf 102}, 139601 (2009).

\bibitem{escudero} C. Escudero, Phys. Rev. Lett. {\bf 100}, 116101
(2008).

\bibitem{escudero2} C. Escudero, Ann. Phys. {\bf 324}, 1796 (2009).

\bibitem{escudero3} C. Escudero, J. Stat. Mech. (in press); arXiv:0901.2733.

\end{thebibliography}
\end{document}